\documentclass[12pt]{article}

\usepackage{color}
\usepackage{a4}
\usepackage{latexsym}
\usepackage{cite}
\usepackage{lineno}

\usepackage{color}
\usepackage{xcolor}

\usepackage{array}
\newcolumntype{T}{>{\ttfamily} c}
\newcolumntype{M}{>{$\displaystyle} c <{$}}

\usepackage{epsfig}
\usepackage{graphicx}
\usepackage{amsmath}
\usepackage{amssymb}
\usepackage{multirow}
\allowdisplaybreaks

\usepackage{pslatex}
\usepackage[latin1]{inputenc}
\usepackage[T1]{fontenc}

\usepackage{color,colordvi}
\def\colour4colour#1{\Blue{#1}}

\newcommand{\gsim}{\raisebox{-0.7mm}{$\:\:\stackrel{>}{{\scriptstyle
 \sim}}\:\: $} }

\newcommand{\beq}{\begin{equation}}
\newcommand{\eeq}{\end{equation}}
\newcommand{\bea}{\begin{eqnarray}}
\newcommand{\eea}{\end{eqnarray}}
\newcommand{\nn}{\nonumber}
\newcommand{\MSb}{$\overline{\mbox{MS}}$}
\newcommand{\ra}{\rightarrow}

\newcommand{\als}{\alpha_{\rm s}}
\newcommand{\ars}{a_{\rm s}}

\newcommand{\hspn}{{\hspace{-3mm}}}
\newcommand{\hspp}{{\hspace{3mm}}}

\def\frct#1#2{\mbox{\small{$\displaystyle\frac{#1}{#2}$}}}

\def\as(#1){{\alpha_{\rm s}^{\,#1}}}
\def\ar(#1){{a_{\rm s}^{\,#1}}}

\def\zr#1{{\zeta_{\:\!#1}^{}}}
\def\mus{{\mu^{\,2}}}

\def\B(#1,#2){{\beta_{#1}^{\,#2}}}

\def\col{\color{black}}

\def\nc{{{n_c}}}
\def\ncs{{{n_{c}^{\,2}}}}

\def\ncf{{{n_{c}^{\,4}}}}

\def\ca{{\col{C^{}_A}}}
\def\cas{{\col{C^{\,2}_A}}}
\def\cat{{\col{C^{\,3}_A}}}
\def\caf{{\col{C^{\,4}_A}}}
\def\cf{{\col{C^{}_F}}}
\def\cfs{{\col{C^{\, 2}_F}}}
\def\cft{{\col{C^{\, 3}_F}}}

\def\nf{{\col{n^{}_{\! f}}}}
\def\nfz{{\col{n^{\,0}_{\! f}}}}
\def\nfo{{\col{n^{\,1}_{\! f}}}}
\def\nfs{{\col{n^{\,2}_{\! f}}}}
\def\nft{{\col{n^{\,3}_{\! f}}}}

\def\dfAAnA{{ \col{{d_{\,A}^{\,abcd}\,d_{\,A}^{\,abcd} / n_a} }}}
\def\dfRAnA{{ \col{{d_{\,R}^{\,abcd}\,d_{\,A}^{\,abcd} / n_a} }}}
\def\dfRRnA{{ \col{{d_{\,R}^{\,abcd}\,d_{\,R}^{\,abcd} / n_a} }}}

\def\dfAAna{{ \col{{d^{\,(4)}_{AA} \over n_a} }}}
\def\dfRAna{{ \col{{d^{\,(4)}_{RA} \over n_a} }}}
\def\dfRRna{{ \col{{d^{\,(4)}_{RR} \over n_a} }}}

\def\xm1{{(1 \! - \! x)}}
\def\xp1{{(1 \! + \! x)}}
\def\LntO{\ln(1\!-\!x)}
\def\Lnt(#1){\ln^{\,#1}(1\!-\!x)}
\def\pqq(#1){p_{\rm{qq}}(#1)}

\def\z#1{{\zeta_{#1}}}

\def\S(#1){{{S}_{#1}}}
\def\Ss(#1,#2){{{S}_{#1,#2}}}
\def\Sss(#1,#2,#3){{{S}_{#1,#2,#3}}}
\def\Ssss(#1,#2,#3,#4){{{S}_{#1,#2,#3,#4}}}
\def\Sssss(#1,#2,#3,#4,#5){{{S}_{#1,#2,#3,#4,#5}}}
\def\Ssssss(#1,#2,#3,#4,#5,#6){{{S}_{#1,#2,#3,#4,#5,#6}}}
\def\Sssssss(#1,#2,#3,#4,#5,#6,#7){{{S}_{#1,#2,#3,#4,#5,#6,#7}}}

\def\Sp(#1,#2){{{S}_{#1}^{\,#2}}}

\def\H(#1){{\rm{H}}_{#1}}
\def\Hh(#1,#2){{\rm{H}}_{#1,#2}}
\def\Hhh(#1,#2,#3){{\rm{H}}_{#1,#2,#3}}
\def\Hhhh(#1,#2,#3,#4){{\rm{H}}_{#1,#2,#3,#4}}
\def\Hhhhh(#1,#2,#3,#4,#5){{\rm{H}}_{#1,#2,#3,#4,#5}}
\def\Hhhhhh(#1,#2,#3,#4,#5,#6){{\rm{H}}_{#1,#2,#3,#4,#5,#6}}

\def\ddelta_{{\delta}}
\def\Dplus(#1){\mathcal{D}_{#1}}
\def\D(#1){{ D_{#1}}}
\def\Dd(#1,#2){{  D_{#1}^{\,#2}}}

\def\dots{..}

\begin{document}
\setlength{\parskip}{0.2cm}
\setlength{\baselineskip}{0.54cm}


\begin{titlepage}
\noindent
ZU-TH 47/24 \hfill October 2024 \\
DESY-24-144 \\
LTH 1384
\vspace{0.3cm}
\begin{center}
{\LARGE \bf Four-loop splitting functions in QCD \\[2mm]
-- The gluon-gluon case --}\\
\vspace{1.7cm}
\large
G.~Falcioni$^{\, a,b}$, F.~Herzog$^{\, c}$, S. Moch$^{\, d}$,
A. Pelloni$^{\, e}$ and A. Vogt$^{\, f}$\\

\vspace{1.0cm}
\normalsize
{\it $^a$Dipartimento di Fisica, Universit\`{a} di Torino,
  Via Pietro Giuria 1, 10125 Torino, Italy}\\
\vspace{1mm}
{\it $^b$ Physik-Institut, Universit\"{a}t Z\"{u}rich, 
  Winterthurerstrasse 190, 8057 Z\"{u}rich, Switzerland}\\
\vspace{5mm}
{\it $^c$Higgs Centre for Theoretical Physics, School of Physics and Astronomy\\
  The University of Edinburgh, Edinburgh EH9 3FD, Scotland, UK}\\
\vspace{5mm}
{\it $^d$II.~Institute for Theoretical Physics, Hamburg University\\
\vspace{0.5mm}
Luruper Chaussee 149, D-22761 Hamburg, Germany}\\
\vspace{4mm}
{\it $^e$Institute for Theoretical Physics, 
  ETH Z\"{u}rich, 8093 Z\"{u}rich, Switzerland} \\
\vspace{4mm}
{\it $^f$Department of Mathematical Sciences, University of Liverpool\\
\vspace{0.5mm}
Liverpool L69 3BX, United Kingdom}\\
\vspace{1.7cm}
{\large \bf Abstract}
\vspace{-0.2cm}
\end{center}
We have computed the even-$N$ moments $N\!\leq\!20$ of the gluon-gluon 
splitting function $P_{\rm gg}$ at the fourth order of perturbative QCD via 
the renormalization of off-shell operator matrix elements.
Our results, derived analytically for a general compact simple gauge group, 
agree with all results obtained for this function so far, in particular with 
the lowest five moments obtained via structure functions in deep-inelastic
scattering.
Using our new moments and all available endpoint constraints, we construct 
improved approximations for the four-loop $P_{\rm gg}(x)$ that should be 
sufficient for a wide range of collider-physics applications.  
The~N$^3$LO contributions to the scale derivative of the gluon distribution,
resulting from these and the corresponding quark-to-gluon splitting 
functions, amount to 1\% or less at $x \gsim 10^{-4}$ at a
standard reference scale with $\als = 0.2$.
\vspace*{0.5cm}
\end{titlepage}


In the next decade, measurements at the Large Hadron Collider (LHC) and the
future Electron-Ion Collider (EIC) will reduce the experimental uncertainties
of many processes towards 1\% 
\mbox{\cite{Dainese:2019rgk,AbdulKhalek:2021gbh}}.
For important observables this accuracy requires theoretical calculations at
the next-to-next-to-next-to-leading order (N$^3$LO) of perturbative QCD.
Fully consistent calculations at this order need, besides the corresponding 
partonic cross sections, the four-loop contributions to the splitting functions
governing the scale dependence (evolution) of the parton distribution functions
(PDFs).
So~far the splitting functions are fully known only to the third order
in the strong coupling $\als$ \cite{Moch:2004pa,Vogt:2004mw}.

The PDFs and splitting functions can be decomposed into flavour non-singlet
(ns) and singlet quantities. Here we focus on the evolution of the singlet 
quark and gluon distributions,
\beq
\label{eq:sgEvol}
 \frac{d}{d \ln\mus} \;
 \Big( \begin{array}{c}
         \! q_{\rm s}^{} \!\! \\ \!g\! \end{array} \Big)
 \: = \: \left( \begin{array}{cc}
         \! P_{\rm qq} & P_{\rm qg} \!\!\! \\
         \! P_{\rm gq} & P_{\rm gg} \!\!\! \end{array} \right)
 \otimes
 \Big( \begin{array}{c}
         \! q_{\rm s}^{}\!\! \\ \!g\!  \end{array} \Big)
 \quad \mbox{with} \quad
 P_{\,\rm ik}^{}(x,\als)
    \,=\, \sum_{n=0} \ar(n+1)\,P_{\,\rm ik}^{\,(n)}(x)
 \: ,
\eeq
where $P_{\rm qq}$ is the sum $P_{\rm ns} + P_{\rm ps}$ (pure singlet), 
$\otimes$ denotes the Mellin convolution in the momentum variable $x$,
and the expansion parameter is chosen as $\ars = \als/(4\:\!\pi)$.
The N$^3$LO contributions $P_{\rm ns}^{\,(3)}$ have been addressed in 
ref.~\cite{Moch:2017uml}; additional results 
beyond the fully known limit of a large number of colours $n_c$ 
will be presented elsewhere.
The exact $x$-dependence of $P_{\,\rm ik}^{\,(3)}$ in eq.~(\ref{eq:sgEvol})
is known only for all leading $\nft$ contributions in the limit of a large 
number $\nf$ of light flavours \cite{Davies:2016jie} and for the $\nfs$ parts 
of $P_{\rm ps}^{\,(3)}$ and $P_{\rm gq}^{\,(3)}$
\cite{Gehrmann:2023cqm,Falcioni:2023tzp}. By themselves, however, these
results are not relevant for phenomenological analyses -- cf.~the relative
size of the contributions in eq.~(\ref{eq:ggg3-num}) below.

Exact expressions for the whole matrix in eq.~(\ref{eq:sgEvol}) do not appear 
to be imminent. It is possible, however, to construct approximations by 
combining the computation of the first even-$N$ Mellin moments,
\beq
\label{eq:Mtrf}
  \gamma_{\,\rm ik}^{\,(n)}(N)
  \;=\; -\int_0^1 \!dx\:x^{\,N-1}\,P_{\,\rm ik}^{\,(n)}(x)
\; ,
\eeq
with known constraints in the high-energy (small-$x$) and threshold 
(large-$x$)limits.
This was done successfully at N$^2$LO in ref.~\cite{vanNeerven:2000wp}.
In the present case the moments $N \leq 10$ have been computed in 
refs.~\cite{Moch:2021qrk,Moch:2023tdj}, which is just about sufficient to 
build first meaningful approximations \cite{Moch:2023tdj}.
With the present means these N$^3$LO calculations, performed via inclusive 
deep-inelastic scattering (DIS), cannot be extended to higher values of $N$.  

We have therefore undertaken to determine $\gamma_{\,\rm ik}^{\,\,(3)}(N)$ in 
eq.~(\ref{eq:Mtrf}) at $N \leq 20$ in 
the conceptually more involved, but computationally simpler framework of the
operator-product expansion, and to provide approximations of 
$P_{\,\rm ik}^{\,(3)}(x)$ that should be sufficient for most collider-physics
applications. The results for $\gamma^{\,(3)}_{\,\rm ps}$, 
$\gamma^{\,(3)}_{\,\rm qg}$ and $\gamma^{\,(3)}_{\,\rm gq}$ have been published
in refs.~\cite{Falcioni:2023luc,Falcioni:2023vqq,Falcioni:2024xyt}.
In this letter, we complete this programme by presenting
$\gamma^{\,(3)}_{\,\rm gg}(N)$ at $N \leq 20$ and providing accurate 
approximations of $P_{\rm gg}^{\,(3)}(x)$.
 
Here and in refs.~\cite{Falcioni:2023luc,Falcioni:2023vqq,Falcioni:2024xyt},  
we have determined the moments (\ref{eq:Mtrf}) as the anomalous dimensions of 
the gluon and quark twist-2 operators $O_{\rm g}$ and $O_{\rm q}$ 
via the renormalization of the off-shell operator matrix elements (OMEs) 
$\rm A_{\rm ij} 
= \langle \,\rm{j}(p) | O_{\:\!\rm i} | \,\rm{j}(p)\rangle$, $\rm i,j=g,q$. 
The determination of $\gamma_{\,\rm gg}^{\,(3)}$ requires the OMEs $\rm A_{gg}$ 
to four loops and $\rm A_{qg}$ to three loops. The latter contribution enters 
via the mixing of $O_{\rm q}$ and $O_{\rm g}$.
In addition, $O_{\rm g}$ mixes also with unphysical `alien' operators,
see refs.~\cite{Falcioni:2022fdm,Gehrmann:2023ksf} and references therein,
which we organise in classes of increasing complexity~\cite{Falcioni:2022fdm}.

The operators of the classes $O_i^{\,I}$ and $O_i^{\,II}$, for $i=A,c$, and 
$O_i^{\,III}$, for $i=A_1,A_2,c$, were already needed for the determination of
$\gamma^{\,(3)}_{\,\rm gq}$; the associated renormalisation constants were 
obtained in refs.~\cite{Falcioni:2024xyt,Falcioni:2024xav}. 
In principle, the calculation of $\gamma_{\rm gg}$ involves more complicated 
operators, listed in ref.~\cite{Falcioni:2022fdm}. 
However, we find that most of these contributions vanish upon contracting the 
OMEs with the physical projector \cite{Dixon:1974ss,Hamberg:1991qt}
\beq
  P_{\mu\,\nu}(N) \;=\; (\Delta \cdot q)^N 
  \left[ g_{\mu\,\nu}(\Delta \cdot q)^2
  - (\Delta_\mu \, q_\nu + \Delta_\nu \, q_\mu)\,\Delta \cdot q 
  + \Delta_\mu\Delta_\nu \,q^2 \right]
\; ,
\eeq
where $\Delta_\mu$ is a lightlike vector and $q_\mu$ the momentum of the 
external gluon. Hence the only new operator that contributes to the 
renormalization of the physical projection of $\rm A_{\rm gg}$ is
\beq
  \mathcal{O}_{c_2}^{\,III} \;=\; 
  -d_4^{\,abcd} \left(\partial\bar{c}^{\,a}\right)
  \sum_{i+j+k^{} \,=\, N-4} \, 3\:\! \kappa^{(2)}_{ijk} \, 
  \big( \partial^{\,i}A^b \big) 
  \big( \partial^{\,j}A^c \big) 
  \big( \partial^{\,k+1}c^{\,d\,}\big)
\: .
\eeq
Here $\partial \,=\, \Delta^{\mu} \, \partial_{\mu}$, 
$d_4^{\,abcd} \,=\, 1/24 \left[\, \text{Tr}\left(T_A^aT_A^bT_A^cT_A^d\right)
+ \text{permutations }\{a,b,c,d\}\right]$ 
with $(T_A^{a})_{bc}^{} = if^{bac}$, where $f^{\,bac}$ are the structure 
constant of the gauge group, and $A$ and $c$ denote the gluon and ghost fields.
All mixing constants $\kappa^{\,(2)}_{\,ijk}$ are given to the relevant 
perturbative order in ref.~\cite{Falcioni:2024xav}.
The~OMEs $A_{\rm ig}$ are required to three loops for every alien operator $i$ 
in the classes listed above.

The Feynman diagrams for the calculation of the OMEs, for both physical and
alien operators, were generated using {\sc Qgraf} \cite{Nogueira:1991ex}.
They were processed by a {\sc Form} 
\cite{Vermaseren:2000nd,Kuipers:2012rf,Ruijl:2017dtg} program that classifies
them according to their colour factors \cite{vanRitbergen:1998pn} and 
topologies. We have computed the resulting propagator-type integrals for even 
$N \leq 20$ using an optimized in-house version of {\sc Forcer} 
\cite{Ruijl:2017cxj}. 
The \MSb\ renormalization of the results yield the N$^3$LO anomalous dimensions 
$\gamma_{\,\rm gg}^{\,(3)}(N)$ for any compact simple gauge group in terms 
of rational numbers and the values $\zr3$, $\zr4$, $\zr5$ of Riemann's 
$\zeta$-function. 
Our new results for $N \geq 12$ are given in app.~\ref{sec:appA}.
The numerical values for QCD~read
\bea
\gamma_{\,\rm gg}^{\,(3)}(N\!=\!2) \; & =\! & 
    \phantom{39876.1233 - \:\:} 
    654.462778 \,\nf
  - 245.610620 \,\nfs
  + 0.92499097 \,\nft
\: , \nn \\[-0.5mm]
\gamma_{\,\rm gg}^{\,(3)}(N\!=\!4) \; & =\! & 
    39876.1233
  - 10103.4511 \,\nf
  + 437.098848 \,\nfs
  + 12.9555655 \,\nft
\: , \nn \\[-0.5mm]
\gamma_{\,\rm gg}^{\,(3)}(N\!=\!6) \; & =\! & 
    53563.8435
  - 14339.1310 \,\nf
  + 652.777331 \,\nfs
  + 16.6541037 \,\nft
\: , \nn \\[-0.5mm]
\gamma_{\,\rm gg}^{\,(3)}(N\!=\!8) \; & =\! & 
    62279.7438
  - 17150.6968 \,\nf
  + 785.880613 \,\nfs
  + 18.9331031 \,\nft
\: , \nn \\[-0.5mm]
\gamma_{\,\rm gg}^{\,(3)}(N\!=\!10) &\! =\! & 
    68958.7532
  - 19307.3854 \,\nf
  + 883.929802 \,\nfs
  + 20.6112832 \,\nft
\: , \nn \\[-0.5mm]
\gamma_{\,\rm gg}^{\,(3)}(N\!=\!12) &\! =\! & 
    74473.0024
  - 21076.0320 \,\nf
  + 962.264417 \,\nfs
  + 21.9511603 \,\nft
\: , \nn \\[-0.5mm]
\gamma_{\,\rm gg}^{\,(3)}(N\!=\!14) &\! =\! & 
    79209.0111
  - 22583.5268 \,\nf
  + 1027.80706 \,\nfs
  + 23.0713754 \,\nft
\: , \nn \\[-0.5mm]
\gamma_{\,\rm gg}^{\,(3)}(N\!=\!16) &\! =\! & 
    83378.4014
  - 23901.3437 \,\nf
  + 1084.30677 \,\nfs
  + 24.0362925 \,\nft
\: , \nn \\[-0.5mm]
\gamma_{\,\rm gg}^{\,(3)}(N\!=\!18) &\! =\! & 
    87112.4096
  - 25074.2309 \,\nf
  + 1134.04028 \,\nfs
  + 24.8850403 \,\nft
\: , \nn \\[-0.5mm]
\gamma_{\,\rm gg}^{\,(3)}(N\!=\!20) &\! =\! & 
    90499.2530
  - 26132.2983 \,\nf
  + 1178.50283 \,\nfs
  + 25.6433278 \,\nft
\: . 
\label{eq:ggg3-num}
\eea

As required by the momentum sum rule,
the result for $\gamma_{\,\rm gg}^{\,(3)}(N\!=\!2)$ is minus that for 
$\gamma_{\,\rm qg}^{\,(3)}(N\!=\!2)$ in ref.~\cite{Falcioni:2023vqq}.
At $N \leq 10$ the results agree with refs.~\cite{Moch:2021qrk,Moch:2023tdj}.
The coefficients of $\nft$ in eq.~(\ref{eq:ggg3-num}) agree with the all-$N$ 
results, also computed via DIS, in eq.~(3.14) of ref.~\cite{Davies:2016jie}.

\newpage
 
Besides the leading large-$\nf$ contributions, all-$N$ expressions for 
$\gamma_{\,\rm gg}^{\,(3)}$ have been obtained until now only for the 
$\zr4$ part, in eq.~(12) of ref.~\cite{Davies:2017hyl}, and the
$\zr5$ coefficients of the quartic group invariants \cite{Moch:2018wjh}.
Using our results to $N=20$, we have now been able to determine at all $N$ 
the $\zr5$ coefficients for all colour factors and the $\zr3$ terms of the 
$\nfs$ contributions.

The all-$N$ expressions of the anomalous dimensions up to N$^3$LO -- as far 
as they are known now -- can be expressed in terms of powers of simple
denominators $1/(N+a) \equiv D_a$ and harmonic sums $S_{\vec{w}}(N)$ 
\cite{Vermaseren:1998uu}, see also ref.~\cite {Blumlein:1998if}. 
Below the denominators will occur mostly, but not exclusively, in the
$N \ra -N-1$ reciprocity invariant combinations
\beq
 \eta \:=\: \D(0)  - \D(1) \:\equiv\: \frct{1}{N}   - \frct{1}{N+1}
 \;\; , \quad 
 \nu  \:=\: \D(-1) - \D(2) \:\equiv\: \frct{1}{N-1} - \frct{1}{N+2} 
\;\; .
\eeq
Note that the results in refs.~\cite{Davies:2016jie} and \cite{Davies:2017hyl}
are given for a slightly different notation with $\nu = \D(-1)\,\D(2)$. 

The complete all-$N$ form of the $\zr5$ contribution to 
$\gamma_{\,\rm gg}^{\,(3)}$ reads
\def\col{\color{blue}}
\bea
\label{eq:ggg3z5N} \mbox{\hspn}
 &&  \gamma_{gg}(N)\big|_{\zeta_5} \;=\; 
       \caf  \,\*  \big(
       - 8/27 \,\* N \* (N+1)
       - 12016/27
       - 1120/3 \,\* \nu
       - 640/3 \,\* \nu^2
       + 3008/9 \,\* \eta
       + 640 \,\* \eta^2
\nn \\[0.5mm] & & \mbox{\hspp}
       + 1760/3 \,\* S_{1}
       + 640/3 \,\* S_{1} \* \left\{2\*\nu - 2\*\eta - S_{1}\right\}
       \big)
     + \dfAAnA  \*  \big(
       - 32/9\, \* N \* (N+1)
       - 48064/9
\nn \\[1mm] & & \mbox{\hspp}
       - 4480 \,\* \nu
       - 2560 \,\* \nu^2
       + 12032/3\, \* \eta
       + 7680 \,\* \eta^2
       + 7040 \,\* S_{1}
       + 2560 \,\* S_{1} \* \left\{2\*\nu - 2\*\eta - S_{1}\right\}
       \big)
\nn \\[1mm] & & \mbox{\hspp}
    + \nf\, \* \cat \*  \big(
        16/27 \,\* N \* (N+1)
      + 272/27
      + 1280/9 \,\* \nu
      + 2384/9 \,\* \eta
      - 12160/3 \,\* \eta^2
      + 160/3 \,\* S_{1}
      \big)
\nn \\[1mm] & & \mbox{\hspp}
     + \nf\, \* \cas \* \cf  \*  \big(
         160
       - 5440/3 \,\* \nu
       + 3280 \,\* \eta
       + 14720 \,\* \eta^2
       + 160 \,\* S_{1}
       \big)
     + \nf\, \* \ca \* \cfs  \*  \big(
         2560 \,\* \nu
\nn \\[1mm] & & \mbox{\hspp}
       - 6480 \,\* \eta
       - 20960 \,\* \eta^2
       - 320 \,\* S_{1}
       \big)
       + \nf\, \* \cft \*  \big(
       - 2560/3 \,\* \nu
       + 2720 \,\* \eta
       + 10560 \,\* \eta^2
       \big)
\nn \\[1mm] & & \mbox{\hspp}
     + \nf \,\* \dfRAnA  \*  \big(
         128/9 \,\* N \* (N+1)
       + 36736/9
       + 2560/3 \,\* \nu
       - 15488/3 \,\* \eta
       + 6400 \,\* \eta^2
\nn \\[1mm] & & \mbox{\hspp}
       - 2560 \,\* S_{1}
       \big)
     + \nfs \* \cas  \*  \big(
       - 8/27 \,\* N \* (N+1)
       - 136/27
       + 368/9 \,\* \eta
       - 640/3 \,\* \eta^2
       \big)
\nn \\[1mm] & & \mbox{\hspp}
     + \nfs \* \dfRRnA  \*  \big(
       - 128/9 \,\* N \* (N+1)
       - 2176/9
       + 5888/3 \,\* \eta
       - 10240 \,\* \eta^2
       \big)
\: .
\eea
\def\col{\color{black}}
\noindent
Here and below we suppress the argument $N$ for brevity. 
The corresponding $\nfs$ terms with $\zr3$ are  
\def\col{\color{blue}}
\bea
\label{eq:ggg3z3N} \mbox{\hspn}
%
&& \gamma_{gg}(N)\big|_{\zeta_3} \;=\;
       \nfs \* \cas \*  \big(
       - 32/135 \,\* N \* (N+1) \* S_{-2}
       - 104/135
       - 114464/405 \,\* \nu
       - 448/9 \,\* \nu^2
\nn \\[0.5mm] & & \mbox{\hspp}
       + 6364/9 \,\* \eta
       + 6872/9 \,\* \eta^2
       + 192 \,\* \eta^3
       - 544/135 \,\* S_{-2}
       - 448/45 \,\* S_{-2} \,\* \eta
       + 640/9 \,\* S_{1}
\nn \\[1mm] & & \mbox{\hspp}
       + 3680/27 \,\* S_{1} \* \nu
       - 4016/9 \,\* S_{1} \* \eta
       - 2624/9 \,\* S_{1} \* \eta^2
       + 256/3 \,\* S_{1} \* \eta^3
       + 64/3 \,\* \Sp(1,2) \,\* \eta \* \left\{ 1 - 4\*\eta \right\}
       \big)
\nn \\[1mm] & & \mbox{\hspp}
     + \nfs \,\* \ca \* \cf \*  \big(
         32/27 \,\* \delta(N-2) 
       - 160/9
       + 3904/9 \,\* \nu
       + 256/3 \,\* \nu^2
       - 11152/9 \,\* \eta
       - 5728/3 \,\* \eta^2
\nn \\[0.5mm] & & \mbox{\hspp}
       - 1472/3 \,\* \eta^3
       + 32/3 \,\* \left\{
         7 \,\*D_{-1}
       - 7 \,\* D_{0}
       - 5 \,\* D_{0}^2
       - 4 \,\* D_{0}^3 \right\}
       - 640/9 \,\* S_{1}
       - 2432/9 \,\* S_{1} \* \nu
       + 768 \,\* S_{1} \* \eta
\nn \\[0.5mm] & & \mbox{\hspp}
       + 2560/3 \,\* S_{1} \* \eta^2
       \big)
     + \nfs \* \cfs \*  \big(
       	 176/9
       - 2240/27 \,\* \nu
       - 256/9 \,\* \nu^2
       + 480 \,\* \eta
       + 7264/9 \,\* \eta^2
       + 448/3 \,\* \eta^3
\nn \\[0.5mm] & & \mbox{\hspp}
       + 256/3 \* \left\{
       - 2 \,\* D_{-1}
       + 2 \,\* D_{0}
       + D_{0}^2
       + 2 \,\* D_{0}^3 \right\}
       + 1280/9 \,\* S_{1} \* \nu
       - 448 \,\* S_{1} \* \eta
       - 1408/3 \,\* S_{1} \* \eta^2
       \big) 
\nn \\[1mm] & & \mbox{\hspp}
     + \nfs \* \dfRRnA  \*  \big(
       - 512/45 \,\* N \* (N+1) \* S_{-2}
       - 7424/45
       - 76544/135 \,\* \nu
       + 4224 \,\* \eta
\nn \\[1mm] & & \mbox{\hspp}
       + 3840 \,\* \eta^2
       - 8704/45 \,\* S_{-2}
       - 7168/15 \,\* S_{-2} \,\* \eta
       + 512/9 \,\* S_{1} \* \nu
       - 9728/3 \,\* S_{1} \* \eta
       + 13312/3 \,\* S_{1} \* \eta^2
\nn \\[1mm] & & \mbox{\hspp}
       + 4096 \,\* S_{1} \* \eta^3
       + 1024 \,\* \Sp(1,2) \,\* \eta \* \left\{ 1 - 4\*\eta \right\}
       \big)
\; .
\eea
\def\col{\color{black}}
 
The above expressions include two structures that do not occur in the 
three-loop anomalous dimensions \cite{Moch:2004pa,Vogt:2004mw}, but enter the 
coefficient functions for inclusive DIS at this order in $\als$
\cite{Vermaseren:2005qc}.
These~are the terms with $N(N\!+\!1) \{ \zr5,\, \zr3 S_{-2} \}$, which occur 
in all non-$C_F$ contributions, and the $\zr3\, \delta(N\!-\!2)$ term in the 
$\nfs C_A C_F$ part.
The latter is also present in $\gamma_{\,\rm ns}^{\,(3)}$
\cite{Gehrmann:2023iah}, $\gamma_{\,\rm ps}^{\,(3)}$ 
\cite{Gehrmann:2023cqm} and in $\gamma_{\,\rm gq}^{\,(3)}$ 
\cite{Falcioni:2023tzp}, where it arises from the non-$\zeta$ contribution 
$(N\!-\!2)^{-1}\, S_{-2}(N\!-\!2)$, corresponding to
$\,x^{\:\!-2}\, {\rm H}_{-1,0}(x) = 
 \,x^{\:\!-2}\, [ \,\ln(x) \ln(1+x) + \mbox{Li}_2(-x) ]$
in $x$-space, see refs.~\cite{Remiddi:1999ew,Moch:1999eb}.

As already noted in ref.~\cite{Moch:2017uml}, the $\zr5\, N(N\!+\!1)$ terms
suggest contributions $N(N\!+\!1)\,f(N)$ with
\beq
\label{eq:fNfct}
     f(N) \;=\;
         5 \* \zr5
       + 4 \:\!\* \zr3 \:\!\* \S(-2)
       - 2 \* \S(-5)
       - 4 \:\!\* \Ss(-2,-3)
       + 8 \* \Sss(-2,-2,1)
       + 4 \:\!\* \Ss(3,-2)
       - 4 \:\!\* \Ss(4,1)
       + 2 \* \S(5)
\eeq
which were first encountered and discussed in ref.~\cite{Vermaseren:2005qc}. 
Indeed, the corresponding coefficients in the $\nfs$ parts of
eq.~(\ref{eq:ggg3z5N}) and in eq.~(\ref{eq:ggg3z3N}) are consistent with
eq.~(\ref{eq:fNfct}).
The $N(N\!+\!1)$ contributions to eq.~(\ref{eq:ggg3z5N}) vanish, separately 
for the quadratic and quartic group invariants, for the choice of $\nf$ and 
colour factors that leads to an ${\cal N} \!=\! 1$ supersymmetric theory, 
see eq.~(2.8) of ref.~\cite{Ruijl:2016pkm}. 

With eq.~(\ref{eq:ggg3z5N}), all $\zr5$ contributions to 
$\gamma_{\,\rm ik}^{\,(3)}(N)$ are now completely known. Their combination 
$  \left. ( \gamma_{\,\rm qq} + \gamma_{\,\rm gq}
          - \gamma_{\,\rm qg} - \gamma_{\,\rm gg} ) \right|_\z5 $
vanishes in the same manner for the above ${\cal N} \!=\! 1$ case.
Beyond the results in eq.~(\ref{eq:ggg3z3N}), we have been able to determine
the all-$N$ form so far only for the combination of the $\zr3\, \dfAAnA$ and 
$\zr3\, \nf\,\dfRAnA$ contributions that enters the super\-symmetric limit. 
Its $N(N\!+\!1)\,\S(-2)$ coefficient also agrees with the presence of the 
function (\ref{eq:fNfct}).
 
The above partial all-$N$ results are of theoretical interest but not 
relevant to N$^3$LO analyses at the LHC. 
Until the complete functions $P_{\rm ik}^{\,(3)}(x)$ become known, 
such analyses will have to rely on approximations based on the available 
moments and information about the large-$x$ and small-$x$ limits.
With eq.~(\ref{eq:ggg3-num}) we are now in the position to improve upon
the $N \leq 10$ based approximations of ref.~\cite{Moch:2023tdj}, thus
putting $P_{\rm gg}^{\,(3)}$ on the same footing as
$P_{\rm ps}^{\,(3)}$, $P_{\rm qg}^{\,(3)}$ and $P_{\rm gq}^{\,(3)}$ 
in refs.~\cite{Falcioni:2023luc,Falcioni:2023vqq,Falcioni:2024xyt}.

The large-$x$ behaviour of $P_{\,\rm gg}^{\,(3)}(x)$ has been addressed 
in detail in ref.~\cite{Moch:2023tdj}. It can be written as
\beq
\label{eq:xto1}
  P_{\,\rm gg,\,x\ra 1\,}^{\,(3)}(x) \:\: = \:\;
        A_{\rm g}^{(4)} / {\xm1_+}
  \,+\, B_{\rm g}^{\,(4)} \, \delta \xm1
  \,+\, C_{\rm g}^{\,(4)} \, \LntO
  \,-\, A_{\rm g}^{(4)} + D_{\rm g}^{\,(4)}
\eeq
up to terms that vanish for $x\!\ra\!1$.
Here $A_{\rm g}^{(4)}$ denotes the four-loop (lightlike) cusp anomalous 
dimension~\cite{Henn:2019swt,vonManteuffel:2020vjv},
$B_{\rm g}^{\,(4)}$ is the corresponding virtual anomalous dimension 
\cite{Das:2019btv,Das:2020adl},
and the coefficients $C_{\rm g}^{\,(4)}$ and $D_{\rm g}^{\,(4)}$ are 
functions of lower-order quantities \cite{Dokshitzer:2005bf,Moch:2017uml}.
With the exception of a few $\nfz$ and $\nfo$ contributions to 
$B_{\rm g}^{\,(4)}$, the coefficients in eq.~(\ref{eq:xto1}) are completely 
known, see eqs.~(15), (16) and (20) of ref.~\cite{Moch:2023tdj}. 
Their rounded numerical values for QCD are
\bea
  A_{\rm g}^{\,(4)} &\!=\!& 
        40880.330         \,-\, 11714.246\,\*\nf 
  \,+\, 440.04876\,\*\nfs \,+\, 7.3627750\,\*\nft 
\, ,
\nn \\[-0.5mm]
  B_{\rm g}^{\,(4)} &\!=\!& 
        68587.64 \pm 0.2  \,-\, 18143.983\,\*\nf 
  \,+\, 423.81135\,\*\nfs \,+\, 0.90672154\,\*\nft
\, ,
\nn \\[-0.5mm]
  C_{\rm g}^{\,(4)} &\!=\!& 
        85814.120 \,-\, 13880.515\,\*\nf \,+\, 135.11111\,\*\nfs
\, ,
\nn \\[-0.5mm]
\label{ABCD4gNum}
  D_{\rm g}^{\,(4)} &\!=\!&
        54482.808 \,-\, 4341.1337\,\*\nf \,-\, 21.133333\,\*\nfs 
\; .
\eea
The uncertainty of the $\nfo$ contribution to $B_{\rm g}^{\,(4)}$ is now 
negligible for all practical purposes; that of the $\nfz$, although less
than $10^{-5}$ of its value, is not irrelevant in the present context.

Up to this small uncertainty, the unknown part of $P_{\rm gg}^{\,(3)}(x)$
vanishes as $x \!\ra\! 1$, i.e., its behaviour is the same as that for
the pure-singlet splitting function in ref.~\cite{Falcioni:2023luc}: 
at large $x$, the $\xm1 \ln^{\,\ell} \xm1$ have been determined for 
$\ell = 3,\,4$ \cite{Soar:2009yh}; the coefficient for $\ell = 1,\,2$ are 
unknown.
The dominant (BFKL) small-$x$ terms $x^{\,-1} \ln^{\,m\!} x$ are known 
for $m = 3$ \cite{Jaroszewicz:1982gr,Catani:1989sg} and $m = 2$
\cite{Fadin:1998py,Ciafaloni:1998gs,Ciafaloni:2005cg,Ciafaloni:2006yk},
but not for $m = 1$ and $m = 0$.
The sub-dominant small-$x$ terms of the form $\ln^{\,n\!} x$ have been 
obtained for $n = 4,\,5,\,6$ \cite{Davies:2022ofz}; the remaining three
coefficients are unknown. 

Consequently, the approximations for $P_{\rm gg}^{\,(3)}$ can be 
built in basically the same manner as those for $P_{\rm ps}^{\,(3)}$ 
in ref.~\cite{Falcioni:2023luc}:
the seven unknown endpoint functions above are combined with 10 
choices of a two-parameter rational function that vanishes as 
$x\!\ra 1$ and 8 choices of a non-rational interpolating function.
In this manner we have built 80 trial functions for both `extreme'
values of $B_{\rm g}^{\,(4)}$ in eq.~(\ref{ABCD4gNum}), of which we
choose two representatives that reflect the remaining uncertainties.

\begin{figure}[p]
\vspace{-4mm}
\centerline{\epsfig{file=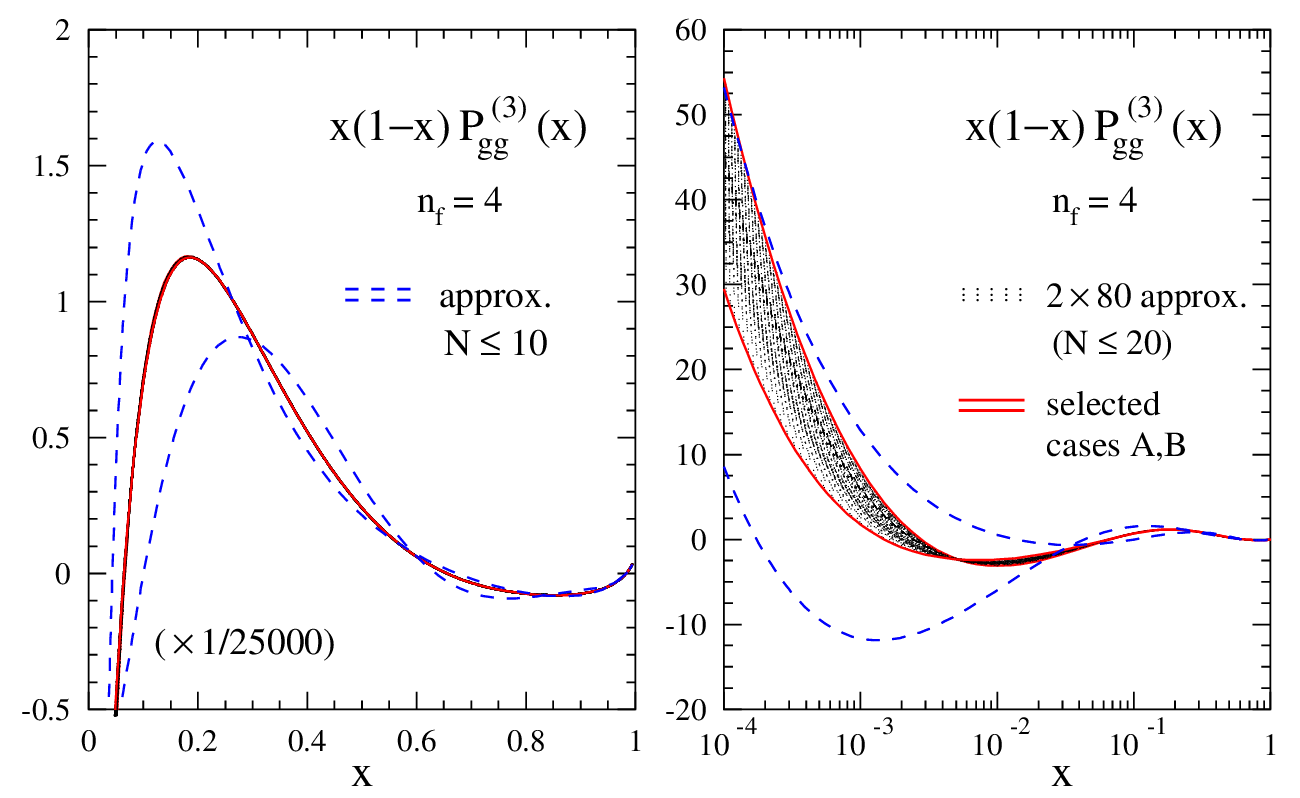,width=16.0cm,angle=0}}
\vspace{-3mm}
\caption{\label{fig:pgq3ab} \small
Two sets of 80 trial functions for the four-loop (N$^3$LO)
contribution to the gluon-gluon splitting function at $\nf = 4$.
The two cases selected for eq.~(\ref{eq:Pgg3A3-nf4}) are shown by
the solid (red) lines. Also shown, by the dashed (blue) lines,
are the selected approximations of ref.~\cite{Moch:2023tdj} based
on only the first 5 even moments.}
\end{figure}

\begin{figure}[p]
\vspace{-2mm}
\centerline{\epsfig{file=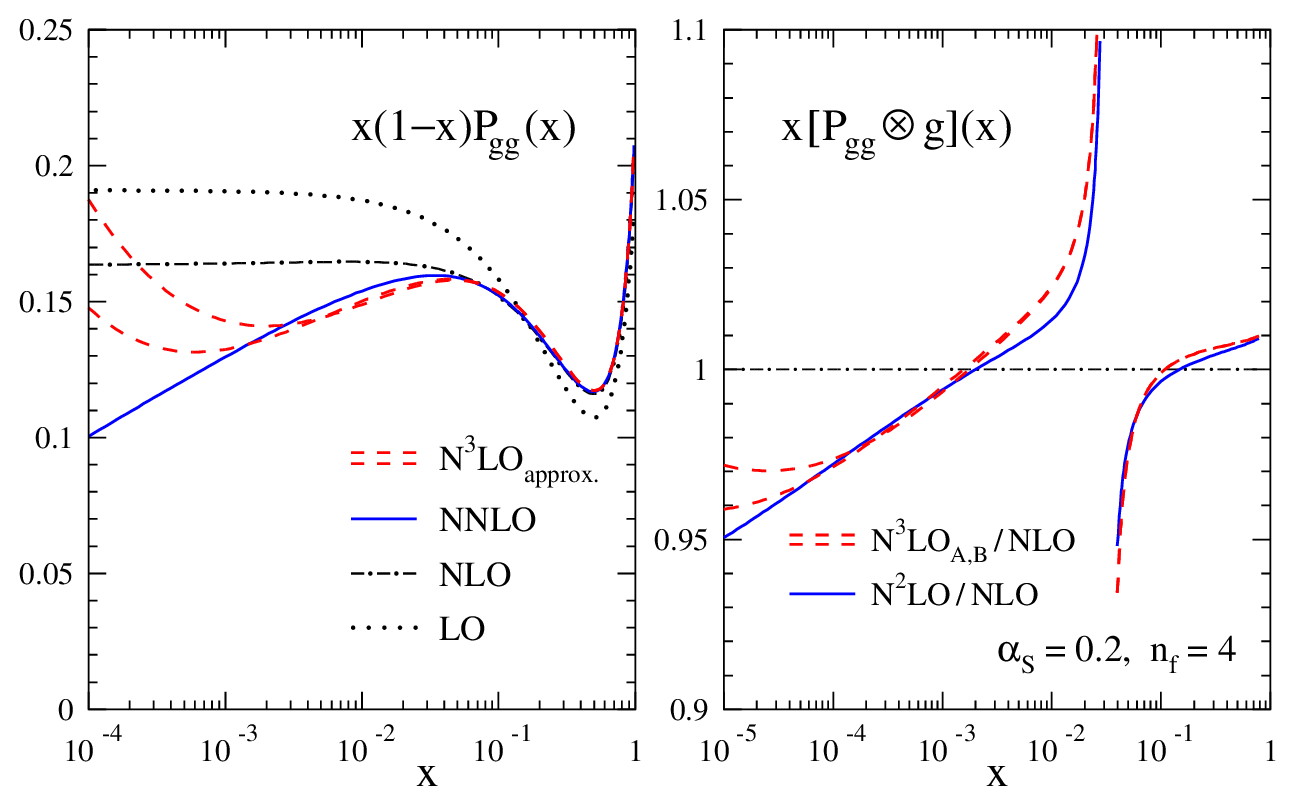,width=16.0cm,angle=0}}
\vspace{-2mm}
\caption{\label{fig:pggn3lo} \small
Left: the perturbative expansion of the splitting functions
$P_{\rm gg}$ to N$^3$LO for $\nf = 4$ and $\als = 0.2$, using
eq.~(\ref{eq:Pgg3A3-nf4}) for the four-loop contribution.
Right: the resulting N$^2$LO and N$^3$LO convolutions with the 
reference gluon distribution in eq.~(\ref{eq:qg-shape}) normalized 
to the NLO result which changes sign at about $x = 0.3$.}
\end{figure}

This procedure is illustrated in fig.~\ref{fig:pgq3ab} for $\nf = 4$.
The 80 trial functions are shown as dotted lines. In the left panel their 
spread is hardly visible, in striking contrast to the uncertainty band, 
shown by the dashed lines, of the previous approximations based on
only five moments \cite{Moch:2023tdj}.
At small-$x$ (right panel), $P_{\rm gg}^{\,(3)}$ is now well constrained at
about $x >\! 10^{-3}$, which represents an improvement by about one order of
magnitude in $x$.
Consequently, the approximation uncertainty is now hardly visible at 
$x \gsim 3\cdot 10^{-3}$ for the total N$^3$LO splitting function $P_{\rm gg}$ 
shown in the left part of fig.~\ref{fig:pggn3lo}.


The selected approximations are built up as
 
\vspace*{-8mm}
\bea
\label{eq:Pgg-def}
  P_{\rm gg,\,A/B}^{\,(3)}(\nf,x) \:=\:
  p_{{\rm gg},0}^{\,[\nf]}(x) \,+\, p_{\rm gg,\,A/B}^{\,[\nf]}(x)
\; ,
\eea
 
\vspace*{-5mm}
\noindent
where $p_{{\rm gg},0}^{\,[\nf]}(x)$ collects the known endpoint contributions
discussed above.
Using the abbreviations $x_1^{} = 1\!-\!x$, $L_1=\ln \xm1$, $L_0 = \ln x$,
and as in eq.~(\ref{ABCD4gNum}) rounding to eight significant figures, we have
\bea
\label{eq:Pgg30-nf}
{\lefteqn{ 
 p_{{\rm gg},0}^{\,[\nf]}(x) \; = \; \mbox{}
   - 8308.6173\,\*L_0^3/x
 \,-\, \left( 106911.99 + 996.38304\,\*\nf \right)\*L_0^2/x
}}
\nn \\[-0.5mm] &&  \mbox{}
 \,+\, \big( 144 - 27.786008 \,\* \nf + 0.79012346 \,\* \nfs \,\big)\* L_0^6
 \,-\, \big( 144 + 162.08066 \,\* \nf - 14.380247 \,\* \nfs \,\big)\* L_0^5
\nn \\ &&  \mbox{}
 \,+\, \big( 26165.784 - 3344.7551  \* \nf
   + 91.522635 \,\* \nfs - 0.19753086 \,\* \nft \,\big)\* L_0^4
\nn \\ &&  \mbox{}
 \,+\, \big( 247.55054 \,\* \nf - 40.559671 \,\* \,\nfs 
       + 1.5802469 \,\* \nft \,\big)\* x_1\*L_1^3
\nn \\[-0.5mm] &&  \mbox{}
 \,+\, \left( 56.460905 \,\* \nf - 3.6213992 \,\* \nfs \,\right)\* x_1\*L_1^4
 \:+\: P_{\rm gg,\,x\ra 1\,}^{\,(3)-/+}(x) 
\eea
with $P_{\rm gg,\,x\ra 1\,}^{\,(3)}(x)$ given by eq.~(\ref{eq:xto1}).
The additional superscript {\small $-/+$} indicates that $-0.2$ in 
eq.~(\ref{ABCD4gNum}) is used for the approximations A, and $+0.2$ for the
approximations B in
\bea
\label{eq:Pgg3A3-nf3}
 p_{\rm gg,\,A}^{\,[3]}(x) &\!=\!\!& \mbox{}
            - 421311 \,\* x_1\*L_0/x
            - 325557 \,\* x_1/x
            + 1679790 \,\* x_1
            - 1456863 \,\* x_1\*x
            + 3246307 \,\* x_1\*L_0
\nn \\[-1mm] && \mbox{}
            + 2026324 \,\* L_0^{2}
            + 549188 \,\* L_0^{3}
            +   8337 \,\* x_1\*L_1
            +  26718 \,\* x_1\*L_1^3
            -  27049 \,\* x_1^{2}\*L_1^3
\, ,
\nn\\
 p_{\rm gg,\,B}^{\,[3]}(x) &\!=\!\!& \mbox{}
            - 700113 \,\* x_1\*L_0/x
            - 2300581 \,\* x_1/x
            + 896407 \,\* x_1\*(1+2 \*x)
            - 162733 \,\* x_1\*x^{2}
\nn \\[-1mm] && \mbox{}
            - 2661862 \,\* x_1\*L_0
            + 196759 \,\* L_0^{2}
            - 260607 \,\* L_0^{3}
            +  84068 \,\* x_1\*L_1
            + 346318 \,\* x_1\*L_1^2
\nn \\[-0.5mm] && \mbox{}
            + 315725 \,\* L_0\*L_1^2
\, ,
\\[0mm]
\label{eq:Pgg3A3-nf4}
 p_{\rm gg,\,A}^{\,[4]}(x) &\!=\!\!& \mbox{}
            - 437084 \,\* x_1\*L_0/x
            - 361570 \,\* x_1/x
            + 1696070 \,\* x_1
            - 1457385 \,\* x_1\*x
            + 3195104 \,\* x_1\*L_0
\nn \\[-1mm] && \mbox{}
            + 2009021 \,\* L_0^{2}
            + 544380 \,\* L_0^{3}
            +  9938  \,\* x_1\*L_1
            +  24376 \,\* x_1\*L_1^2
            -  22143 \,\* x_1^{2}\*L_1^3
\, ,
\nn\\[0.5mm]
 p_{\rm gg,\,B}^{\,[4]}(x) &\!=\!& 
            - 706649 \,\* x_1\*L_0/x
            - 2274637 \,\* x_1/x
            + 836544 \,\* x_1\*(1+2 \*x)
            - 199929 \,\* x_1\*x^{2}
\nn \\[-1mm] && \mbox{}
            - 2683760 \,\* x_1\*L_0
            + 168802 \,\* L_0^{2}
            - 250799 \,\* L_0^{3}
            +  36967 \,\* x_1\*L_1
            +  24530 \,\* x_1\*L_1^2
\nn \\[-0.5mm] && \mbox{}
            -  71470 \,\* x_1^{2}\*L_1^2
\, ,
\\[1mm]
\label{eq:Pgg3A3-nf5}
 p_{\rm gg,\,A}^{\,[5]}(x) &\!=\!& 
            - 439426 \,\* x_1\*L_0/x
            - 293679 \,\* x_1/x
            + 1916281 \,\* x_1
            - 1615883 \,\* x_1\*x
            + 3648786 \,\* x_1\*L_0
\nn \\[-1mm] && \mbox{}
            + 2166231 \,\* L_0^{2}
            + 594588 \,\* L_0^{3}
            +  50406 \,\* x_1\*L_1
            +  24692 \,\* x_1\*L_1^2
            + 174067 \,\* x_1^{2}\*L_1
\, ,
\nn\\[0.5mm]
 p_{\rm gg,\,B}^{\,[5]}(x) &\!=\!& 
            - 705978 \,\*  x_1\*L_0/x
            - 2192234 \,\* x_1/x
            + 1730508 \,\* x_1\*x
            + 353143 \* x_1\*(2-x^{2})
\nn \\[-1mm] && \mbox{}
            - 2602682 \,\* x_1\*L_0
            + 178960 \,\* L_0^{2}
            - 218133 \,\* L_0^{3}
            +   2285 \,\* x_1\*L_1
            +  19295 \,\* x_1\*L_1^2
\nn \\[-0.5mm] && \mbox{}
            -  13719 \,\* x_1^{2}\*L_1^2
\, .
\eea
The error bands above lead to precise numerical predictions for 
$\gamma_{\,\rm gg}^{\,(3)}$ at higher values of $N$ such as
\bea
\label{eq:gqg3N22appr}
  - \gamma_{\,\rm gg}^{\,(3)}(N\!=\!22) \;=\; 
  23990.457275 (20)\, ,\quad
  6396.872080  (20)\, ,\quad
  -8123.349380 (20)\, 
\eea
for $\nf\,=\,3,4,5$, where the brackets indicate a conservative uncertainty 
of the last two digits.

In order to illustrate the size and uncertainties of the present N$^3$LO
contributions to the evolution of the gluon PDF, we adopt the reference point 
of ref.~\cite{Vogt:2004mw} with $\nf = 4$, $\als (\mu_{0}^{\,2}) \ =\, 0.2$ and 
\bea
  xq_{\rm s}^{}(x,\mu_{0}^{\,2}) &\! = \!&
  0.6\: x^{\, -0.3} (1-x)^{3.5}\, \left(1 + 5.0\: x^{\, 0.8\,}\right)
\: ,
\nn \\[0.5mm]
  xg(x,\mu_{0}^{\,2}) &\! = \!&
  1.6\: x^{\, -0.3} (1-x)^{4.5}\, \left(1 - 0.6\: x^{\, 0.3\,}\right)
\; .
\label{eq:qg-shape}
\eea
The resulting N$^2$LO and N$^3$LO contributions to the convolution 
$[P_{\rm gg} \otimes g](x)$ are shown in the right part of 
fig.~\ref{fig:pggn3lo}. 
In order to render these small effects visible, we have normalized these 
corrections to the NLO results, which however leads to an artificial 
singularity slightly above $x = 0.3$ due to a sign change at NLO. 
Except at the smallest values of $x$, the N$^3$LO corrections are below $1\%$. 
Their uncertainties are negligible at $x >\! 10^{-4}$. Even at $x = 10^{-5}$, 
the  N$^3$LO effect amounts to only about 1.5\% with an uncertainty of less 
than $\pm 1\%$.

\begin{figure}[t]
\vspace*{-2mm}
\centerline{\epsfig{file=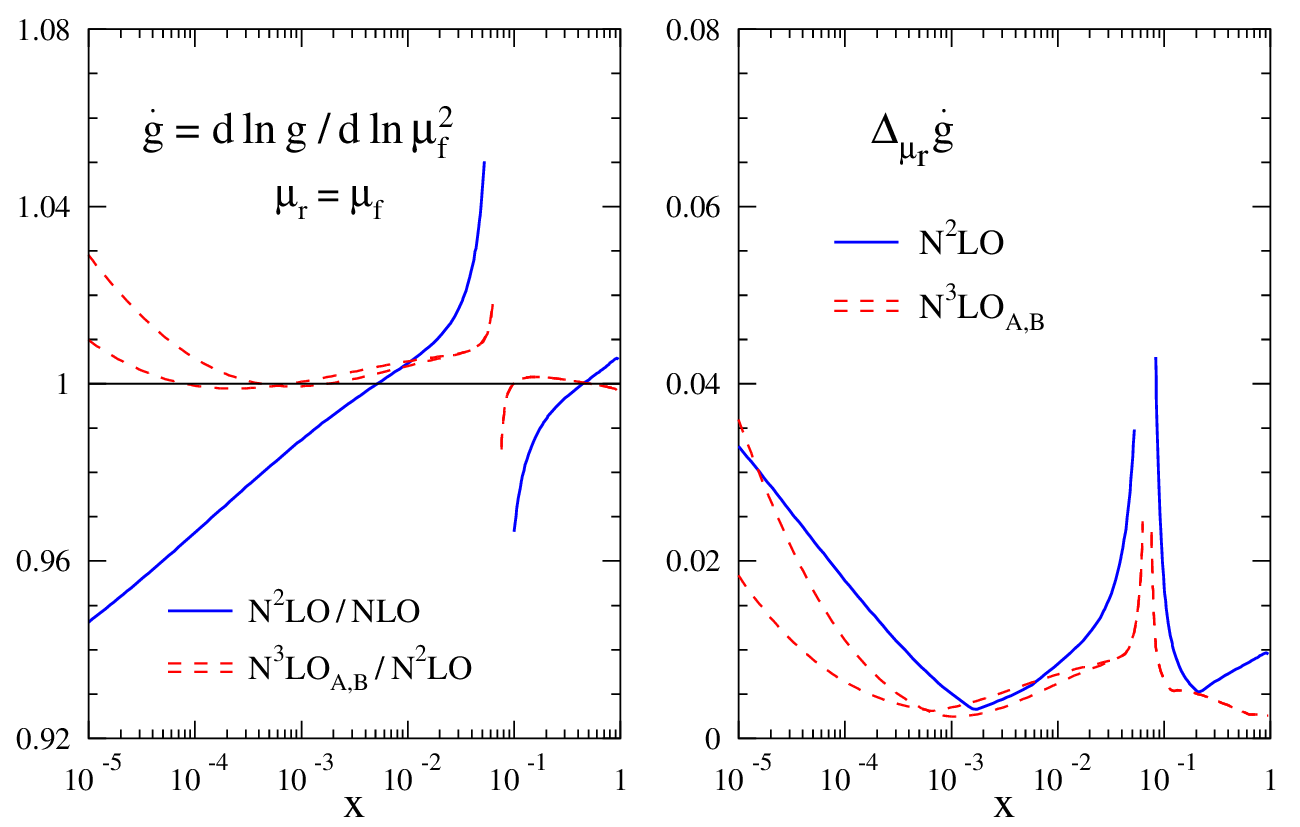,width=15.5cm,angle=0}}
\vspace*{-2mm}
\caption{\label{fig:dgln3lo} \small
Left: the relative N$^2$LO and N$^3$LO corrections to the scale
derivative of the gluon PDF $g$ at the reference point with
$\nf=4$ and $\als(\mu_{0}^{\,2}) = 0.2$.
Right: the renormalization-scale uncertainties of these results,
as estimated using eq.~(\ref{eq:mur-var}).
Note that $dg / d\ln \mu^{\,2}$ changes sign close to
$x=0.07$, which leads to (irrelevant)  singularities in both the
relative corrections and the relative scale uncertainty in eq.~%
(\ref{eq:mur-var}).}
\vspace*{-3mm}
\end{figure}

Combining our above results with those for $P_{\rm gq}^{\,(3)}$ in 
ref.~\cite{Falcioni:2024xyt}, in a manner that the uncertainties add up at 
small $x$, we are finally able to evaluate the N$^3$LO contributions to the 
scale derivative (\ref{eq:sgEvol}) of the gluon PDF at the above reference 
point.
The relative size of the N$^2$LO and  N$^3$LO contributions to
$\dot{g} \,\equiv\, dg / d\ln \mu^{\,2}$ is shown in the left panel of 
fig.~\ref{fig:dgln3lo}. Both the N$^3$LO corrections and their uncertainties
are below 1\% at $x > 10^{-4}$, reaching $(2 \pm 1)\%$ at $x = 10^{-5}$.

So far we have identified the renormalization scale $\mu_{\:\!\rm r}^{}$ 
with the factorization scale $\mu_{\rm f}^{} \equiv \mu$. 
The expansion in eq.~(\ref{eq:sgEvol}) is readily expanded to 
$\mu_{\:\!\rm r}^{} \neq \mu_{\rm f}^{}$, see, e.g., eq.~(2.9) of
ref.~\cite{vanNeerven:2001pe} for the expression to order $\as(5)$.
The scale stability of $\dot{g}$ is illustrated in the right panel of 
fig.~\ref{fig:dgln3lo} by the quantity
\beq
\label{eq:mur-var}
 \Delta_{\,\mu_{\:\!\rm r}^{}} \,\dot{g} \:\: \equiv \:\:
 \frac{
 \max\, [ \,\dot{g}(x,\mu_{\rm r}^{\,2}
   = \lambda\, \mu_{\rm f}^{\,2})]
 \,-\, \min\, [ \,\dot{g} (x,\mu_{\rm r}^{\,2}
   = \lambda\, \mu_{\rm f}^{\,2})] }
 { 2 |\, {\rm average}\, [ \,\dot{g}(x, \mu_r^{\,2}
   = \lambda\, \mu_{\rm f}^{\,2})]\, | }
\eeq
for the conventional range $\lambda =1/4 \,\dots\, 4$.
At $x >\! 10^{-4}$ this N$^3$LO scale uncertainty is below 1\%, except close
to the sign change of $\dot{g}$ at $x \simeq 0.7$. It reaches $(3 \pm 1)\%$ at
$x = 10^{-5}$. This is very similar to the result for the quark PDF in fig.~3
of ref.~\cite{Falcioni:2024xyt}, where the improvement on the N$^2$LO 
uncertainty is much more impressive since the latter is much larger than in 
the present gluon case.

\pagebreak

With this article, we have completed the computation of the moments to $N=20$ 
of the four-loop (N$^3$LO) splitting functions for the evolution of the flavour
singlet PDFs in the \MSb~scheme.
We have combined these moments with knowledge on the $x \!\ra 0$ and 
$x \!\ra\! 1$ limits to construct approximate expressions for 
$P_{\,\rm ik}^{\,(3)}(x)$ that should be sufficient for most high-scale 
applications. 
For example, the N$^3$LO effects on the scale dependence $d\!f / d\ln\mus$, 
$f = q_{\rm s}^{},\, g$ are below 1\% with an 
irrelevant approximation error down to $x \gsim\! 10^{-4}$ at a standard 
reference point with $\mus \,\simeq\, 25\ldots 50$ GeV$^2$,
as shown in the left parts of fig.~3 of ref.~\cite{Falcioni:2023vqq} and 
fig.~3 above. 
The remaining large uncertainties of $P_{\,\rm ik}^{\,(3)}(x)$ for $x \ra 0$
would shrink dramatically if all small-$x$ logarithms 
$x^{-1} \ln^{\,\ell\!} x$ were calculated.

%
\vspace*{-2mm}
\subsection*{Acknowledgements}
\vspace*{-3mm}
This work has been supported by the UKRI FLF Mr/S03479x/1;
the STFC Consolidated Grants ST/X000494/1 and ST/T000988/1;
the EU 
Marie Sklodowska-Curie grant 101104792; 
the DFG through the Research Unit FOR 2926,
project number 40824754, and DFG grant MO~1801/4-2, and 
the ERC Advanced Grant 101095857. 

{\small
\addtolength{\baselineskip}{-2.2mm}
\providecommand{\href}[2]{#2}\begingroup\raggedright\endgroup
}

\vspace{-5mm}
\appendix
%
%
\renewcommand{\theequation}{\ref{sec:appA}.\arabic{equation}}
\setcounter{equation}{0}
\section{Mellin moments of $P^{\,(3)}_{\,\rm gg}$}
\label{sec:appA}

\vspace{-2mm}
The four-loop anomalous dimensions $\gamma^{\,(3)}_{\,\rm gg}(N)$ at even 
$2\leq N \leq 10$ have been already been obtained in 
refs.~\cite{Moch:2021qrk,Moch:2023tdj} 
for a general compact simple gauge group.
Here we report the corresponding expressions for $12\leq N\leq 20$.
The numerical values in QCD have been given in eq.~(\ref{eq:ggg3-num}) above.


The quadratic Casimir invariants in SU$(\nc)$ are $\ca = \nc$ and 
$\cf = (\ncs-1)/(2\nc)$.
The quartic group invariants are products of two symmetrized traces of four 
generators $T_r^a$, 
\beq
\label{eq:d4def}
  d_{r}^{\,abcd} \; =\; \frct{1}{6}\: {\rm Tr} \, ( \, 
   T_{r}^{a\,} T_{r}^{b\,} T_{r}^{c\,} T_{r}^{d\,}
   + \,\mbox{ five $bcd$ permutations}\, ) 
\; , 
\eeq
in the fundamental ($R$) or adjoint ($A$) representation,
which leads to ($n_a = \ncs - 1$)
\beq
\label{eq:quarticsSUN}
\dfAAna \;=\;
  \frac{1}{24}\: \ncs ( \ncs + 36 ) 
\;\,, \quad
  \dfRAna \:=\:
  \frac{1}{48}\: \nc ( \ncs + 6 )
\;\,, \quad
\dfRRna \;=\;
\frac{( \ncf - 6\,\ncs + 18 )}{96\,\ncs}
\; .
\eeq
and hence $\dfAAnA = 135/8$, $\dfRAnA = 15/16$ and $\dfRRnA = 5/96$ in QCD.
\def\col{\color{blue}}

{\small{
\bea
%
%
%
&& \hspn \gamma_{\,\rm gg}^{\,(3)}(N=12) 
\,=\, \caf\,\*\Big(\frac{426916808845670890185486559702559}{435010740735016802996640000000}
+\frac{2858733158927526553}{4499203191210000}\,\*\z3
\nn \\[0.5mm] & & \mbox{\hspp}
-\frac{7173493238}{10061415}\,\*\z5\Big)
+\dfAAna\,\*\Big(\frac{5186984525115894967}{3012146177040000}
+\frac{11837685988559}{2112897150}\,\*\z3
-\frac{28693972952}{3353805}\,\*\z5\Big)
\nn \\[0.5mm] & & \mbox{\hspp}
+\nf\,\*\cat\,\*\Big(-\frac{55171996036371050045998366332119}{102830377785469173455616000000}
-\frac{2216417127123463709}{2304353803752000}\,\*\z3
\nn \\[0.5mm] & & \mbox{\hspp}
+\frac{28951453}{53235}\,\*\z4
+\frac{8698336}{31941}\,\*\z5\Big)
+\nf\,\*\cf\,\*\cas\,\*\Big(-\frac{31891030785065733878575519402109}{89119660747406616994867200000}
\nn \\[0.5mm] & & \mbox{\hspp}
+\frac{21063377310621153257}{57032756642862000}\,\*\z3
-\frac{15914749031}{29066310}\,\*\z4
+\frac{25095032}{39039}\,\*\z5\Big)
\nn \\[0.5mm] & & \mbox{\hspp}
+\nf\,\*\cfs\,\*\ca\,\*\Big(\frac{2879456567418214942254486948055493}{12131413819240725738426297600000}
+\frac{2277122121751096111}{3802183776190800}\,\*\z3
\nn \\[0.5mm] & & \mbox{\hspp}
+\frac{18263117221}{3517023510}\,\*\z4
-\frac{115425482}{117117}\,\*\z5\Big)
+\nf\,\*\cft\,\*\Big(-\frac{180913555304484221774208990520433}{12131413819240725738426297600000}
\nn \\[0.5mm] & & \mbox{\hspp}
-\frac{1713716843924069}{475272972023850}\,\*\z3
-\frac{2642589184}{1758511755}\,\*\z4
+\frac{16220}{13013}\,\*\z5\Big)
+\nf\,\*\dfRAna\,\*\Big(\frac{378251047050369859609}{662672158948800000}
\nn \\[0.5mm] & & \mbox{\hspp}
+\frac{476371067828641}{852201850500}\,\*\z3
-\frac{27773168}{16731}\,\*\z5\Big)
+\nfs\,\*\cas\,\*\Big(\frac{43776212234240781257735569}{6277800841603734643200000}
\nn \\[0.5mm] & & \mbox{\hspp}
+\frac{240437709082559}{949596347700}\,\*\z3
-\frac{57902906}{585585}\,\*\z4
-\frac{8620}{169}\,\*\z5\Big)
+\nfs\,\*\cf\*\ca\*\Big(\frac{3059177558618292646984219483}{40397648415720032428992000}
\nn \\[0.5mm] & & \mbox{\hspp}
-\frac{3778891871983}{15826605795}\,\*\z3
+\frac{58210216}{585585}\,\*\z4\Big)
+\nfs\,\*\cfs\,\*\Big(-\frac{28551595906275919031595376597}{2407026551436651932227440000}
\nn \\[0.5mm] & & \mbox{\hspp}
+\frac{317531792566}{15826605795}\,\*\z3
-\frac{61462}{117117}\,\*\z4\Big)
+\nfs\,\*\dfRRna\,\*\Big(\frac{5121825047548437461}{7888954273200000}
+\frac{812056934692}{553377825}\,\*\z3
\nn \\[0.5mm] & & \mbox{\hspp}
-\frac{413760}{169}\,\*\z5\Big)
+\nft\,\*\ca\,\*\Big(-\frac{504853821441288677}{410635847828606400}
+\frac{809864}{110565}\,\*\z3\Big)
\nn \\[0.5mm] & & \mbox{\hspp}
+\nft\,\*\cf\,\*\Big(-\frac{6113543346521554934099}{12331394510293050192000}
-\frac{24964}{351351}\,\*\z3\Big)
\, , \\[3mm]
%
%
%
&&\hspn \gamma_{\,\rm gg}^{\,(3)}(N=14)
\,=\,\caf\,\*\Big(\frac{3530302858272626511166908679609871}{3320926870659095011362048000000}
+\frac{392576018133364531}{527175090288000}\,\*\z3
\nn \\[0.5mm] & & \mbox{\hspp}
-\frac{12182898853}{14407470}\,\*\z5\Big)
+\dfAAna\,\*\Big(\frac{7388026987231247233447}{3568234702032000000}
+\frac{7007627407900661}{1058949045000}\,\*\z3
\nn \\[0.5mm] & & \mbox{\hspp}
-\frac{24365797706}{2401245}\,\*\z5\Big)
+\nf\,\*\cat\,\*\Big(-\frac{113594703594760740723691396139053}{198218775567347268814080000000}
\nn \\[0.5mm] & & \mbox{\hspp}
-\frac{543858408413511329}{524956339908000}\,\*\z3
+\frac{81775856}{143325}\,\*\z4
+\frac{294336388}{945945}\,\*\z5\Big)
\nn \\[0.5mm] & & \mbox{\hspp}
+\nf\,\*\cf\,\*\cas\,\*\Big(-\frac{28644681812359432810528968921761}{73624116639300414130944000000}
+\frac{61123089029529077743}{153549729423090000}\,\*\z3
\nn \\[0.5mm] & & \mbox{\hspp}
-\frac{189827271499}{331080750}\,\*\z4
+\frac{70425496}{105105}\,\*\z5\Big)
+\nf\,\*\cfs\,\*\ca\,\*\Big(\frac{6363232067291046187032334411336813}{25124229803161266322184640000000}
\nn \\[0.5mm] & & \mbox{\hspp}
+\frac{729557540910110873}{1163255525932500}\,\*\z3
+\frac{17085172697}{4304049750}\,\*\z4
-\frac{326322424}{315315}\,\*\z5\Big)
\nn \\[0.5mm] & & \mbox{\hspp}
+\nf\,\*\cft\,\*\Big(-\frac{4166118844338186141796182641273}{245390674260778709382720000000}
-\frac{10174644862216}{3776803655625}\,\*\z3
-\frac{505959889}{430404975}\,\*\z4
\nn \\[0.5mm] & & \mbox{\hspp}
+\frac{2816}{3185}\,\*\z5\Big)
+\nf\,\*\dfRAna\,\*\Big(\frac{16222306083794825031617}{31891097649411000000}
+\frac{21688516374213961}{82024428110625}\,\*\z3
-\frac{399720544}{315315}\,\*\z5\Big)
\nn \\[0.5mm] & & \mbox{\hspp}
+\nfs\,\*\cas\,\*\Big(\frac{86264786812375898336153759}{7150749479341532064000000}
+\frac{2455639519261343}{8948119430250}\,\*\z3
-\frac{163551712}{1576575}\,\*\z4
-\frac{16432}{245}\,\*\z5\Big)
\nn \\[0.5mm] & & \mbox{\hspp}
+\nfs\,\*\cf\,\*\ca\,\*\Big(\frac{4787481590605144814770575829}{59759834934497089392000000}
-\frac{371371561826}{1489863375}\,\*\z3
+\frac{164162696}{1576575}\,\*\z4\Big)
\nn \\[0.5mm] & & \mbox{\hspp}
+\nfs\,\*\cfs\,\*\Big(-\frac{473951339528575686520746571}{39653140472161089523650000}
+\frac{385728471251}{19368223875}\,\*\z3
-\frac{55544}{143325}\,\*\z4\Big)
\nn \\[0.5mm] & & \mbox{\hspp}
+\nft\,\*\ca\,\*\Big(-\frac{49815330459045199}{38655875938680000}
+\frac{718924}{93555}\,\*\z3\Big)
+\nft\,\*\cf\,\*\Big(-\frac{3907253297496328220297}{7545433703850642600000}
\nn \\[0.5mm] & & \mbox{\hspp}
-\frac{22472}{429975}\,\*\z3\Big)
+\nfs\,\*\dfRRna\,\*\Big(\frac{4485052909723377974579}{5315182941568500000}
+\frac{226041368622508}{114719479875}\,\*\z3
-\frac{788736}{245}\,\*\z5\Big)
\, , \\[3mm]
%
%
%
&&\hspn\gamma_{\,\rm gg}^{\,(3)}(N=16)
\,=\,\caf\,\*\Big(\frac{15337182291951433616165281475248927995540971534167}{13492339347924264874246991614296194641920000000}
\nn \\[0.5mm] & & \mbox{\hspp}
+\frac{5529754419565844478534458563}{6520610325382626432960000}\,\*\z3
-\frac{38027329168168}{39093069015}\,\*\z5\Big)
\nn \\[0.5mm] & & \mbox{\hspp}
+\dfAAna\,\*\Big(\frac{17225198856494540525364768331}{7156940604431871052800000}
+\frac{4280088643520652417443}{563503134009816000}\,\*\z3
\nn \\[0.5mm] & & \mbox{\hspp}
-\frac{152109316672672}{13031023005}\,\*\z5\Big)
+\nf\,\*\cat\,\*\Big(-\frac{897432462848236045017850628184670314114341}{1478553186610374665701444214833152000000}
\nn \\[0.5mm] & & \mbox{\hspp}
-\frac{133102211488321772967121}{120260238076679500800}\,\*\z3
+\frac{11241734927}{18935280}\,\*\z4
+\frac{16593150641}{46864818}\,\*\z5\Big)
\nn \\[0.5mm] & & \mbox{\hspp}
+\nf\,\*\cf\,\*\cas\,\*\Big(-\frac{26847608088981105116818871646074617292318363}{64633896443253521100663132819849216000000}
\nn \\[0.5mm] & & \mbox{\hspp}
+\frac{236351484521633019349663}{558351105356011968000}\,\*\z3
-\frac{10549880344093}{17704486800}\,\*\z4
+\frac{10826183435}{15621606}\,\*\z5\Big)
\nn \\[0.5mm] & & \mbox{\hspp}
+\nf\,\*\cfs\,\*\ca\,\*\Big(\frac{3820884848912654785490648965626200547231253}{14363088098500782466814029515522048000000}
\nn \\[0.5mm] & & \mbox{\hspp}
+\frac{74860657727420155363}{114839799538464000}\,\*\z3
+\frac{160275239209}{50988921984}\,\*\z4
-\frac{22442950375}{20828808}\,\*\z5\Big)
\nn \\[0.5mm] & & \mbox{\hspp}
+\nf\,\*\cft\,\*\Big(-\frac{432465293017180154359501484955972282652531}{23503235070274007672968411934490624000000}
-\frac{3753742543711044061}{1802101469680512000}\,\*\z3
\nn \\[0.5mm] & & \mbox{\hspp}
-\frac{1209091491169}{1274723049600}\,\*\z4
+\frac{20639}{31212}\,\*\z5\Big)
+\nf\,\*\dfRAna\,\*\Big(\frac{412776185633813536481909239}{1022420086347410150400000}
\nn \\[0.5mm] & & \mbox{\hspp}
-\frac{18868730583338146213}{140875783502454000}\,\*\z3
-\frac{5575791757}{7810803}\,\*\z5\Big)
\nn \\[0.5mm] & & \mbox{\hspp}
+\nfs\,\*\cas\,\*\Big(\frac{1411423039987918734914010233965753}{80450768545744814319745843200000}
+\frac{679370984698281247}{2296795990769280}\,\*\z3
\nn \\[0.5mm] & & \mbox{\hspp}
-\frac{11241734927}{104144040}\,\*\z4
-\frac{1334035}{15606}\,\*\z5\Big)
+\nfs\,\*\cf\,\*\ca\,\*\Big(\frac{265603320468239150618251481266503761}{3165163093928303123379715031040000}
\nn \\[0.5mm] & & \mbox{\hspp}
-\frac{1482304546167539}{5736253723200}\,\*\z3
+\frac{8454580201}{78108030}\,\*\z4\Big)
\nn \\[0.5mm] & & \mbox{\hspp}
+\nfs\,\*\cfs\,\*\Big(-\frac{11501285202205811759108125497437825419}{956582623942776055065869431603200000}
+\frac{113706461016607}{5736253723200}\,\*\z3
-\frac{93023}{312120}\,\*\z4\Big)
\nn \\[0.5mm] & & \mbox{\hspp}
+\nfs\,\*\dfRRna\,\*\Big(\frac{63517459719119486443848313}{59641171703598925440000}
+\frac{7173470647621241}{2814700969080}\,\*\z3
-\frac{10672280}{2601}\,\*\z5\Big)
\nn \\[0.5mm] & & \mbox{\hspp}
+\nft\,\*\ca\,\*\Big(-\frac{1696185564468333948773}{1265075231715719424000}
+\frac{680336}{85085}\,\*\z3\Big)
\nn \\[0.5mm] & & \mbox{\hspp}
+\nft\,\*\cf\,\*\Big(-\frac{2758634195497061062738263139}{5166668452345535385169920000}
-\frac{18769}{468180}\,\*\z3\Big)
\, \\[3mm]
%
%
%
&&\hspn\gamma_{\,\rm gg}^{\,(3)}(N=18)
\,=\,\caf\,\*\Big(\frac{613821523893530269831391065477998922572407445587423}{509646570159129631368783252529910692372070400000}
\nn \\[0.5mm] & & \mbox{\hspp}
+\frac{2600432772102362174102051929}{2747076348676753152259200}\,\*\z3
-\frac{7132610803355}{6511002498}\,\*\z5\Big)
\nn \\[0.5mm] & & \mbox{\hspp}
+\dfAAna\,\*\Big(\frac{1049053721644320302406316291}{383787218983943174400000}
+\frac{2277808717857396553}{266625552293100}\,\*\z3
-\frac{14265221606710}{1085167083}\,\*\z5\Big)
\nn \\[0.5mm] & & \mbox{\hspp}
+\nf\,\*\cat\,\*\Big(-\frac{50902451110310945082922336631719197853721939}{79640562719441064394469307771805685760000}
\nn \\[0.5mm] & & \mbox{\hspp}
-\frac{932396716084659320721077653}{793111952814766894108800}\,\*\z3
+\frac{46298577401}{75393045}\,\*\z4
+\frac{31516663202}{78567489}\,\*\z5\Big)
\nn \\[0.5mm] & & \mbox{\hspp}
+\nf\,\*\cf\,\*\cas\,\*\Big(-\frac{4472469837937412890257391830178132693234847257}{10207897523166453888656185083449697024000000}
\nn \\[0.5mm] & & \mbox{\hspp}
+\frac{134159922529333641505923173}{300957660666317794639500}\,\*\z3
-\frac{87339087634793}{141814317645}\,\*\z4
+\frac{354751643702}{497594097}\,\*\z5\Big)
\nn \\[0.5mm] & & \mbox{\hspp}
+\nf\,\*\cfs\,\*\ca\,\*\Big(\frac{647979310225413508460477144500358567981204150383}{2342712481566701167446594476651705467008000000}
\nn \\[0.5mm] & & \mbox{\hspp}
+\frac{1418122620537283713547619}{2104599025638585976500}\,\*\z3
+\frac{30822057470944}{12054216999825}\,\*\z4
-\frac{554803942604}{497594097}\,\*\z5\Big)
\nn \\[0.5mm] & & \mbox{\hspp}
+\nf\,\*\cft\,\*\Big(-\frac{70301841538977369648349286711177952437421103729}{3620555653330356349690191463916272085376000000}
\nn \\[0.5mm] & & \mbox{\hspp}
-\frac{26374016931923364628}{15943932012413530125}\,\*\z3
-\frac{9447656272424}{12054216999825}\,\*\z4
+\frac{85184}{165699}\,\*\z5\Big)
\nn \\[0.5mm] & & \mbox{\hspp}
+\nf\,\*\dfRAna\,\*\Big(\frac{580668376095736761345039157}{2241459420069322252800000}
-\frac{28798161501291573012959}{45770542059947304600}\,\*\z3
\nn \\[0.5mm] & & \mbox{\hspp}
-\frac{4582985696}{497594097}\,\*\z5\Big)
+\nfs\,\*\cas\,\*\Big(\frac{7280509142736822801047375373219121409}{310179343928745969019165270487040000}
\nn \\[0.5mm] & & \mbox{\hspp}
+\frac{827826218263460428211}{2606314582834162200}\,\*\z3
-\frac{92597154802}{829323495}\,\*\z4
-\frac{9320800}{87723}\,\*\z5\Big)
\nn \\[0.5mm] & & \mbox{\hspp}
+\nfs\,\*\cf\,\*\ca\,\*\Big(\frac{619976473559417663859944479698070705897}{7103660867653869736912133203564800000}
-\frac{2224097820277294}{8345227153725}\,\*\z3
\nn \\[0.5mm] & & \mbox{\hspp}
+\frac{278379603958}{2487970485}\,\*\z4\Big)
+\nfs\,\*\cfs\,\*\Big(-\frac{29450269754989290176205939653770867886633}{2437887614017962420336532713798394800000}
\nn \\[0.5mm] & & \mbox{\hspp}
+\frac{238194199445648}{12054216999825}\,\*\z3
-\frac{587552}{2485485}\,\*\z4\Big)
+\nfs\,\*\dfRRna\,\*\Big(\frac{508929299629112028606527827567}{387548333729985817509120000}
\nn \\[0.5mm] & & \mbox{\hspp}
+\frac{215307386311880818}{67242378298095}\,\*\z3
-\frac{149132800}{29241}\,\*\z5\Big)
+\nft\,\*\ca\,\*\Big(-\frac{2522300408158699916579371}{1818373558151738283696000}
\nn \\[0.5mm] & & \mbox{\hspp}
+\frac{38224084}{4621617}\,\*\z3\Big)
+\nft\,\*\cf\,\*\Big(-\frac{1204343230800942414809786168123}{2204707477283604289122906780000}
-\frac{236672}{7456455}\,\*\z3\Big)
\, \\[3mm]
%
%
%
&&\hspn\gamma_{\,\rm gg}^{\,(3)}(N=20)
\,=\,\caf\,\*\Big(\frac{514425183460510317418209161756016723035932624964779}{405894542050054968300263664396743855477760000000}
\nn \\[0.5mm] & & \mbox{\hspp}
+\frac{7638681169638394821437322497}{7335686616055454737080000}\,\*\z3
-\frac{31656671189543}{26062046010}\,\*\z5\Big)
\nn \\[0.5mm] & & \mbox{\hspp}
+\dfAAna\,\*\Big(\frac{3889870578491714739384956213923949}{1274320113146361260467488000000}
+\frac{3934478309186512848349}{415641774465918000}\,\*\z3
\nn \\[0.5mm] & & \mbox{\hspp}
-\frac{63313342379086}{4343674335}\,\*\z5\Big)
+\nf\,\*\cat\,\*\Big(-\frac{76969971585375286644444812731996776908979058237}{114836399358229868686405566463915499520000000}
\nn \\[0.5mm] & & \mbox{\hspp}
-\frac{10836873355305051655841337131}{8714711208213966795840000}\,\*\z3
+\frac{9758076254}{15431325}\,\*\z4
+\frac{27649610258}{61108047}\,\*\z5\Big)
\nn \\[0.5mm] & & \mbox{\hspp}
+\nf\,\*\cf\,\*\cas\,\*\Big(-\frac{74982413930823242524026665139426671440525248653}{163641869085477562878127932211079586816000000}
\nn \\[0.5mm] & & \mbox{\hspp}
+\frac{91863486093996885376294187}{197118467804839725144000}\,\*\z3
-\frac{71007716588687}{112031419500}\,\*\z4
+\frac{74400774706}{101846745}\,\*\z5\Big)
\nn \\[0.5mm] & & \mbox{\hspp}
+\nf\,\*\cfs\,\*\ca\,\*\Big(\frac{4439838820552240984417113640877343980811677953361}{15545977563120368473422153560052560747520000000}
\nn \\[0.5mm] & & \mbox{\hspp}
+\frac{7217347848995756680249831}{10405952297389729080000}\,\*\z3
+\frac{7914266258212}{3725044698375}\,\*\z4
-\frac{116965983598}{101846745}\,\*\z5\Big)
\nn \\[0.5mm] & & \mbox{\hspp}
+\nf\,\*\cft\,\*\Big(-\frac{309438164995702284150415680403691064724486193873}{15344081750612311740001086630701228789760000000}
\nn \\[0.5mm] & & \mbox{\hspp}
-\frac{10474380989917400036281}{7804464223042296810000}\,\*\z3
-\frac{9834028074797}{14900178793500}\,\*\z4
+\frac{63236}{153615}\,\*\z5\Big)
\nn \\[0.5mm] & & \mbox{\hspp}
+\nf\,\*\dfRAna\,\*\Big(\frac{67402279701398725909910179689001}{876455462435506387289856000000}
-\frac{44176589943925468760202749}{36311296700891528316000}\,\*\z3
\nn \\[0.5mm] & & \mbox{\hspp}
+\frac{85425608144}{101846745}\,\*\z5\Big)
+\nfs\,\*\cas\,\*\Big(\frac{17645361960130685484820958817420805056989}{590479223355768555565639482023424000000}
\nn \\[0.5mm] & & \mbox{\hspp}
+\frac{846815221459124600897}{2489462272102806000}\,\*\z3
-\frac{19516152508}{169744575}\,\*\z4
-\frac{855884}{6615}\,\*\z5\Big)
\nn \\[0.5mm] & & \mbox{\hspp}
+\nfs\,\*\cf\,\*\ca\,\*\Big(\frac{18088671861812820773497194991710703558327}{200341165067135759924056252829376000000}
-\frac{1833909745247}{6698382075}\,\*\z3
\nn \\[0.5mm] & & \mbox{\hspp}
+\frac{58646289362}{509233725}\,\*\z4\Big)
+\nfs\,\*\cfs\,\*\Big(-\frac{143611841677221812934598916644203681895721}{11842388868412913808844214056136448000000}
\nn \\[0.5mm] & & \mbox{\hspp}
+\frac{24036876714101}{1219105537650}\,\*\z3
-\frac{442678}{2304225}\,\*\z4\Big)
+\nfs\,\*\dfRRna\,\*\Big(\frac{6993864067805233426081908293714947}{4402196754505611627069504000000}
\nn \\[0.5mm] & & \mbox{\hspp}
+\frac{10728928931473261367}{2729673543972375}\,\*\z3
-\frac{13694144}{2205}\,\*\z5\Big)
+\nft\,\*\ca\,\*\Big(-\frac{178413939710515996456301}{124871427568676748960000}
\nn \\[0.5mm] & & \mbox{\hspp}
+\frac{8385068}{984555}\,\*\z3\Big)
+\nft\,\*\cf\,\*\Big(-\frac{12122556111894521113303106571737}{21801854470925127152156803200000}
-\frac{178084}{6912675}\,\*\z3\Big)
\, .
\eea

}}

\vspace*{-1mm}
{\sc Form} files with the results for $\gamma_{\,\rm gg}^{}(N)$ at
even $N \leq 20$ and the partial all-$N$ expressions in the main text
have been deposited at the preprint server {\tt http://arXiv.org}
together with a {\sc Fortran} subroutine of our approximations for the
splitting function $P_{\,\rm gg}^{\,(3)}(x)$ in
eqs.~(\ref{eq:Pgg30-nf}) - (\ref{eq:Pgg3A3-nf5}).
These files are also available from the authors upon request.
\end{document}